\documentclass[twocolumn,showkeys,showpacs,preprintnumbers,prd,superscriptaddress,nofootinbib]{revtex4-1}
\usepackage{graphicx}
\usepackage{epsf}
\usepackage{bm}
\usepackage{amsmath}
\usepackage{amsfonts}
\usepackage{amssymb}
\usepackage{epstopdf}
\usepackage{natbib}
\usepackage{hyperref}
\usepackage{color}
\usepackage{verbatim}
\usepackage{multirow}
\usepackage{bm}
\usepackage{hyperref}
\usepackage{soul}

\begin{document}

\title{Updated Cosmological Constraints from 2D BAO Measurements: A New Compilation and Comparison with DESI DR2}

\author{Miguel A. Sabogal}
\email{miguel.sabogal@ufrgs.br}
\affiliation{Instituto de Física, Universidade Federal do Rio Grande do Sul, 91501-97 Porto Alegre, RS, Brazil}
\affiliation{Observatório Nacional, Rua General José Cristino 77, São Cristóvão, 20921-400 Rio de Janeiro, RJ, Brazil}

\author{Rafael C. Nunes}
\email{rafadcnunes@gmail.com}
\affiliation{Instituto de Física, Universidade Federal do Rio Grande do Sul, 91501-97 Porto Alegre, RS, Brazil}
\affiliation{Divisão de Astrofísica, Instituto Nacional de Pesquisas Espaciais, 
Avenida dos Astronautas 1758, 12227-010 São José dos Campos, SP, Brazil}

\author{Felipe Avila}
\email{felipeavila@on.br}
\affiliation{Observatório Nacional, Rua General José Cristino 77, São Cristóvão, 20921-400 Rio de Janeiro, RJ, Brazil}

\author{Armando Bernui}
\email{bernui@on.br}
\affiliation{Observatório Nacional, 
Rua General José Cristino 77, 
São Cristóvão, 20921-400 Rio de Janeiro, RJ, Brazil}

\begin{abstract}
We investigate and update observational constraints on cosmological parameters within the $\Lambda$CDM and dynamical dark energy frameworks, using a new 
compilation of the transverse (or 2D) BAO data, 
measurements that provide a relatively model-independent estimate of the BAO angular scale at a given redshift. 
Firstly, we assess the consistency of this compilation with CMB-Planck data and recent BAO results from the DESI collaboration. After confirming minimal tension with CMB data, we perform a series of joint analyses combining CMB data with the 2D~BAO compilation, as well as with several recent Type Ia supernova (SNIa) samples. In all cases, we compare the constraining power of the 2D~BAO data with that of DESI~DR2 
data (3D~BAO). 
Our results indicate that combining 2D~BAO with CMB and SNIa data provides observational constraints that are {in agreement with} those obtained using DESI~DR2 data. Although the precision of DESI~DR2 results remains higher, as expected due to the more accurate 3D measurements, the 2D~BAO compilation 
{combined with other probes} 
yields strong constraints. 
For example, in the $\Lambda$CDM context, we find (CMB + 2D~BAO): $H_0 = 68.16^{+0.41}_{-0.37} \,\,\, \text{km s}^{-1}\,\text{Mpc}^{-1}$ and $\Sigma m_{\nu} < 0.081~\mathrm{eV}$ (95\%~CL). 
These results are consistent with analogous analyses using DESI~DR2. 
Several other cases are investigated and discussed in the main text. 
{Our results show that this new 2D~BAO compilation is robust and delivers meaningful cosmological constraints. Parameters from 2D~BAO alone agree with CMB-only results, with no significant tension. Overall, 2D and 3D BAO provide consistent and complementary information when combined with other probes.
}
\end{abstract}



\maketitle

\section{Introduction}
\label{sec:introduction}

The standard cosmological model, the flat \(\Lambda\)CDM, built upon general relativity with a positive cosmological constant and cold dark matter, has successfully explained a broad range of cosmological observations over the past two decades. 
Nevertheless, as astronomical data have become increasingly precise and diverse, a growing number of measurements appear difficult to reconcile within the standard paradigm, placing the \(\Lambda\)CDM cosmology at a potential crossroads. 
The most striking example is the current tension in the determination of the Hubble constant, \(H_0\). 
Under the \(\Lambda\)CDM framework, the analysis of Planck CMB data~\cite{Planck:2018vyg} yields $H_0 = 67.36 \pm 0.54 \; \text{km s}^{-1}\,\text{Mpc}^{-1},$ whereas a model-independent local determination from the SH0ES collaboration, based on long-period Cepheid variables, finds  $H_0 = 73.18 \pm 0.88 \; \text{km s}^{-1}\,\text{Mpc}^{-1}$ \cite{Riess:2025chq}. 
These estimates differ by approximately \(6 \sigma\). The lower value inferred from the Planck-CMB data is, however, in excellent agreement with the joint constraints from Baryon Acoustic Oscillations (BAO) and Big Bang Nucleosynthesis (BBN) \cite{Schoneberg:2024ifp,Schoneberg:2022ggi}, as well as with results from other CMB experiments such as ACT-DR6 and SPT \cite{ACT:2025fju,SPT-3G:2025bzu}. 
Several additional, though less statistically significant, tensions have also been reported in recent years (see \cite{CosmoVerseNetwork:2025alb} for a comprehensive review). 
Taken together, these tensions-- originating from independent and complementary data sets-- have motivated extensive investigations into extensions of the standard \(\Lambda\)CDM model \cite{CosmoVerseNetwork:2025alb}.

On the other hand, recent observations from the Dark Energy Spectroscopic Instrument (DESI) have placed this assumption under increasing scrutiny~\cite{DESI2025}. 
By measuring BAO in an unprecedented sample of more than 14 million galaxies and quasars, DESI second data release (DR2) reports statistically significant deviations from the \(\Lambda\)CDM predictions, with tensions reaching up to \(4.2\sigma\), depending on the sample considered~\cite{DESI2025}. 
These findings may represent a turning point, suggesting that dark energy (DE) might not be well described by a simple cosmological constant. 
Moreover, several independent analyses have confirmed significant deviations from the 
\(\Lambda\)CDM framework and proposed new cosmological tests to further probe it in light of BAO DESI-DR2 samples~\cite{Pedrotti:2025ccw,RoyChoudhury:2025iis,Wolf:2025acj,Dinda:2025hiu,Fazzari:2025lzd,Adam:2025kve,Wu:2025vfs,Paliathanasis:2025kmg,Chaudhary:2025vzy,Paul:2025wix,Arora:2025msq,Camarena:2025upt,Lee:2025pzo,Mishra:2025goj,Hogas:2025ahb,Gialamas:2025pwv,Ozulker:2025ehg,Cline:2025sbt,An:2025vfz,Luciano:2025elo,Manoharan:2025uix,Mukherjee:2025ytj,Bayat:2025xfr,Cheng:2025lod,Ye:2025ark,Liu:2025mub,Silva:2025hxw,Li:2025muv,Toomey:2025xyo,Yang:2025uyv,Paul:2025wix,Wang:2025znm,Bhattacharjee:2025xeb,Qiang:2025cxp,Liu:2025myr,Li:2025ula,vanderWesthuizen:2025iam,Li:2025eqh,Pan:2025upl,Paliathanasis:2025hjw,Ling:2025lmw,Yashiki:2025loj,Liu:2025mub,RoyChoudhury:2025dhe,Kumar:2025etf,Yang:2025ume,Wu:2025wyk,Wolf:2025jed,Pedrotti:2024kpn,Pan:2025qwy,Giare:2025pzu}, with some alternative cosmological models providing a better fit than  \(\Lambda\)CDM to the data by up to \(5\sigma\)~\cite{Scherer:2025esj}.

The growing number of reported tensions with the $\Lambda$CDM model, now also emerging in traditional three-dimensional (3D) BAO measurements, has intensified the need to explore complementary late-time probes capable of testing the standard cosmological model in a minimally correlated and model-independent way. 
In this context, there is an ongoing debate regarding whether the $\theta_{\mathrm{BAO}}$\footnote{The quantity $\theta_{\mathrm{BAO}}$ corresponds to the transverse BAO measurement derived from the two-point angular correlation function between pairs of cosmic tracers, which is used to identify and measure the BAO angular scale at an effective redshift.} measurements themselves show a tension with $\Lambda$CDM (see, e.g.,~\cite{Bernui23,Dwivedi24,Favale24,Zheng25}).

The BAO phenomenon can be studied without assuming a fiducial cosmology by employing the transverse BAO approach, also known as the two-dimensional (2D) BAO. 
An important advantage of this method is that it does not depend on a fiducial cosmology to calculate three-dimensional comoving distances between pairs of cosmic objects, because it analyzes data distributed in spherical shells of redshift thickness \(\Delta z\), and find the BAO signal considering only the angular correlations between pairs. 
The redshift shells used in the analysis must be carefully chosen: if they are too thick, projection effects smooth and shift the BAO signal; if they are too thin, the number density of tracers may become insufficient to yield an adequate signal-to-noise ratio. 

In a deep astronomical survey covering a wide sky area, the observed volume can be divided into several disjoint redshift shells to avoid correlations between adjacent bins. The analysis of these shells provides the angular scale of the 2D BAO signal, \(\theta_{\mathrm{BAO}}(z)\), at each redshift \(z\), or equivalently, measurements of the angular diameter distance \(D_A(z)\), with the comoving sound horizon scale $r_d$ serving as a standard ruler. 
Due to the weakly model-dependent nature of 2D BAO measurements, the set \(\{\theta_{\mathrm{BAO}}(z)\}\) offers a powerful and independent means to test both the standard \(\Lambda\)CDM cosmology and alternative models, as well as to perform cross-comparisons among different observational datasets. Such measurements have been widely employed in recent literature across a broad range of observational tests and analyses~\cite{Nunes:2020hzy,Zheng:2025cgq,Staicova:2025tbi,Dwivedi:2024okk,Giare:2024syw,Liu:2024txl,Wang:2024rxm,Liu:2024dlf, 
Gomez-Valent:2023uof, Akarsu:2023mfb, Wang:2023ghk, Staicova:2021ntm, Benisty:2020otr,Nunes:2020uex, Oliveira25a, Oliveira25b}.

In this work, we investigate observational constraints on cosmological parameters by combining 17 measurements of the 2D BAO scale with state-of-the-art Planck CMB data and several recent Type Ia supernova (SNIa) samples. 
{Our main objectives can be summarized as follows:}

\begin{itemize}
\item Provide an updated 
compilation of 2D BAO measurements for the community. This compilation is minimally consistent with CMB data, allowing joint analyses of CMB and 2D BAO datasets from a general observational perspective. {The main purpose of this compilation is not to extract new cosmological constraints with higher precision, but rather to serve as an independent validation of high-precision 3D BAO results~\citep{DESI2025}. 
Unlike 3D BAO, which rely on full three-dimensional clustering information and require reconstruction procedures, redshift-space distortion modeling, and Alcock–Paczyński rescalings relative to a fiducial cosmology, 2D BAO 
measurements are obtained from the angular correlation function within narrow redshift shells. 
Consequently, 2D BAO are minimally affected by redshift-space distortions, and require little to no Alcock–Paczyński correction. 
This difference in the observational pipeline and modeling assumptions underlines the value of 2D BAO measurements as an independent and systematic consistency check for 3D BAO analyses, despite 
being less precise measurements.}

{It is important to emphasize that 2D BAO measurements do not provide additional cosmological information beyond that already obtained through high-precision 3D BAO analyses. 
Instead, their primary role is to serve as an independent consistency check. Because 2D BAO are derived from the angular correlation function within narrow redshift shells, they are subject to different observational and modeling systematics compared to 3D BAO.}

\item Update observational constraints within the \(\Lambda\)CDM framework and its minimal extensions, such as the CPL parametrization~\cite{Chevallier2000,Linder2002}, by combining 2D BAO and CMB measurements with the most recent SNIa data, including PantheonPlus, DES Y5, and Union 3.0.
    
\item Compare the constraining power of cosmological parameters of this 2D BAO compilation with that of the BAO DESI DR2 collaboration.
\end{itemize}

{These objectives offer new perspectives for conducting cosmological tests and serve as a reference for researchers interested in incorporating 2D BAO measurements into cosmological parameter estimation. 
The primary goal of this work is to present an updated compilation of 2D BAO data, a dataset that is weakly model-dependent and with observational systematics distinct from that of 3D BAO measurements, 
considered, therefore, suitable for an independent consistency check. 
As a secondary goal, we compare the constraining power of cosmological parameters from this 2D BAO compilation with those obtained by the BAO DESI DR2 collaboration. Although 2D BAO measurements are less precise than full 3D BAO data, this comparison allows us to assess the consistency of the results, investigate potential differences, and demonstrate the complementarity of these independent datasets. 
Overall, our results provide a robust and timely update on the role of 2D BAO measurements in a broader cosmological context, serving as a robustness test of cosmological inference rather than a new precision determination.}

This work is organized as follows. In Section~\ref{sec:data_method}, we present a new compilation of 2D BAO measurements, along with the complementary datasets that will be used in our analysis. The theoretical methodology is also briefly described in this section. In Section~\ref{sec:results}, we present and discuss our main results. Finally, in Section~\ref{sec:conclusions}, we summarize our conclusions and outline prospects for future work.

\section{The Data set and Methodology}
\label{sec:data_method}

The large-scale structure of the Universe is well described by the Friedmann–Lemaître–Robertson–Walker (FLRW) metric, which assumes a homogeneous and isotropic spacetime. In comoving coordinates, the line element can be written as
\begin{equation}
ds^2 = -c^2 dt^2 + a^2(t) \left[ \frac{dr^2}{1 - K r^2} + r^2 (d\theta^2 + \sin^2\theta\, d\phi^2) \right],
\end{equation}
where $a(t)$ is the scale factor and $K$ is the spatial Gaussian curvature ($K = 0, > 0, < 0$ for Euclidean, spherical, and hyperbolic spatial geometries, 
respectively).

Assuming that General Relativity accurately describes the dynamics of the cosmic expansion, the evolution of the Hubble parameter $H(z) \equiv \dot{a}/a$ is governed by the Friedmann equation, which can be expressed as
\begin{align}
\frac{H(z)}{H_0} &= 
\Bigg[ (\Omega_{b} + \Omega_{\mathrm{cdm}})(1 + z)^3 + 
\Omega_{\gamma}(1 + z)^4 + \Omega_{K}(1 + z)^2 \nonumber\\
&\quad + \Omega_{\nu} \frac{\rho_{\nu}(z)}{\rho_{\nu,0}} + 
\Omega_{\mathrm{DE}} \frac{\rho_{\mathrm{DE}}(z)}{\rho_{\mathrm{DE},0}} \Bigg]^{1/2} \,.
\label{eq:Friedmann}
\end{align}
Here, $\Omega_b$, $\Omega_c$, $\Omega_{\gamma}$, 
$\Omega_{K}$, $\Omega_{\nu}$, and $\Omega_{\mathrm{DE}}$ represent the present-day fractional energy densities of baryons, cold dark matter, radiation, curvature, 
neutrinos, and dark energy, respectively. 

In the $\Lambda$CDM framework, dark energy is modeled as a cosmological constant, $\Lambda$, with an energy density $c^2 \rho_{\mathrm{DE}}$ that does not vary 
with time or spatial position. More generally, if dark energy is characterized by an equation-of-state parameter
\begin{equation}
w(z) \equiv \frac{P(z)}{c^2 \rho_{\mathrm{DE}}(z)} \,,
\end{equation}
where $P(z)$ denotes its pressure, its energy density evolves according to
\begin{equation}
\frac{\rho_{\mathrm{DE}}(z)}{\rho_{\mathrm{DE},0}} =
\exp \left[ 3 \int_0^z \frac{1 + w(z')}{1 + z'} \, dz' \right] \,.
\label{eq:rhoDE}
\end{equation}

For a constant $w$, Eq.~\eqref{eq:rhoDE} simplifies to $\rho_{\mathrm{DE}}(z)/\rho_{\mathrm{DE},0} = (1 + z)^{3(1 + w)}$, with $w = -1$ corresponding to a true cosmological constant. 
A widely used parametrization of $w$ as a function of the scale factor $a(z) = (1 + z)^{-1}$ is the CPL model~\cite{Chevallier2000,Linder2002} 
\begin{equation}
w(a) = w_0 + w_a (1 - a) \,,
\label{eq:w0wa}
\end{equation}
which evolves from $w \simeq w_0 + w_a$ at early times to $w_0$ today. This form captures the behavior of many physically motivated dark energy models \cite{dePutter:2008wt}, though more intricate evolution patterns are possible. Cosmological models that combine cold dark matter with the parametrization in Eq.~\eqref{eq:w0wa} are typically referred to as $w_0 w_a$CDM. In this work, we adopt the same standard notation.

This parametrization provides a consistent evolution of $w$ across the entire history of the Universe, preventing non-physical divergences both at early times ($z \to \infty$, where $w \to w_0 + w_a$) and in the distant future ($z \to -1$, where $w \to w_0$). 
A notable property of the CPL model is that it reduces to the standard $\Lambda$CDM scenario when $w_0 = -1$ and $w_a = 0$, allowing for straightforward and systematic tests of the cosmological constant model within a broader dynamical dark energy context.

More broadly, a variety of other two-parameter dark energy models based on $w_0$ and $w_a$ 
have been proposed in the literature \cite{Barboza:2008rh, Dimakis:2016mip, Jassal:2005qc, Efstathiou:1999tm}. 
Nevertheless, from the viewpoint of observational constraints, these alternatives produce 
essentially identical predictions \cite{Giare:2024gpk,Xu:2025nsn,Lee:2025pzo,Barua:2025ypw}. Therefore, for the objectives of this study, 
the CPL parametrization is fully adequate, as it encapsulates the key phenomenology while avoiding superfluous parametric complexity.

The theoretical scenarios considered in this work were implemented within the \texttt{CLASS} Boltzmann solver~\cite{Blas:2011rf}, while parameter inference was performed using the \texttt{MontePython} sampler~\cite{Brinckmann:2018cvx, Audren:2012wb}. We employed Monte Carlo analyses based on Markov Chain Monte Carlo (MCMC) techniques, {to obtain state-of-the-art constraints on our main baseline cosmological parameters}.

{The main cosmological parameters sampled in this study encompass both the standard $\Lambda$CDM quantities and extensions relevant for dark energy phenomenology. Specifically, we consider the physical baryon density}, 
\begin{equation}
\omega_{\mathrm{b}} = \Omega_{\mathrm{b}} h^2,
\end{equation} 
{which governs the baryonic contribution to the total matter content and affects the relative heights of acoustic peaks in the CMB power spectrum. The physical cold dark matter density,}
\begin{equation}
\omega_{\mathrm{c}} = \Omega_{\mathrm{c}} h^2,
\end{equation} 
{controls the growth of large-scale structure and the overall shape of the matter power spectrum. 
The angular size of the sound horizon at recombination, $\theta_{\mathrm{s}}$, determines the characteristic scale of the CMB acoustic peaks and is tightly constrained by observations. 
The reionization optical depth, $\tau$, impacts the amplitude of the large-scale CMB polarization and encodes information about the epoch of reionization.}  

{We also vary the parameters describing the primordial perturbations: the scalar amplitude of primordial fluctuations, $A_{\mathrm{s}}$, which sets the overall normalization of the initial power spectrum, and the scalar spectral index, $n_{\mathrm{s}}$, which governs the scale dependence of these perturbations. In addition, we include the sum of the neutrino masses, $\Sigma m_{\nu}$, which affects both the expansion history and the growth of structures at late times.}

{Beyond the standard $\Lambda$CDM parameters, we explore the dark energy sector by varying $w_0$ and $w_a$ within the framework of the $w_0w_a$CDM model defined above, allowing for a time-dependent equation of state parameterized as $w(a) = w_0 + w_a (1-a)$.}

{In addition, we verified the convergence of all Markov Chain Monte Carlo chains using the Gelman–Rubin diagnostic~\cite{Gelman_1992}, requiring that}
\begin{equation}
R - 1 \leq 10^{-2}.
\end{equation}
{This criterion ensures that the posterior distributions are robust and not biased by insufficient sampling.}

\begin{table}
	\begin{center}
		\renewcommand{\arraystretch}{1.4}
		\begin{tabular}{|c@{\hspace{1 cm}}|@{\hspace{1 cm}} c|}
			\hline
			\textbf{Parameter}           & \textbf{Prior}\\
			\hline\hline

$\omega_{b}$                 & $[0.01,\,1.0]$ \\ 
$\omega_{\rm cdm}$           & $[0.01,\,1.0]$ \\ 
$\tau_{\rm reio}$            & $[0.004,\,0.8]$ \\ 
$n_s$                        & $[0.2,\,2.0]$ \\ 
$\ln(10^{10} A_s)$           & $[1.0,\,5.0]$ \\ 
$100\,\theta_s$              & $[0.5,\,2.0]$ \\ \hline
$w_0$                     & $[-3,\,1]$ \\ 
$w_a$                & $[-3,\,2]$ \\ \hline
$\sum m_\nu\;[\mathrm{eV}]$  & $[0,\,10]$ \\ 		
			\hline
		\end{tabular}
	\end{center}
	\caption{Flat (uniform) priors are imposed on all free parameters used in the statistical analyses. Parameters related to the neutrino sector are varied only in the extended scenarios discussed in Section~\ref{neutrino_results}.}
	\label{tab:priors}
\end{table}

{In Table~\ref{tab:priors}, we summarize the uniform (flat) priors adopted in our statistical analyses. These choices are motivated by the considerations discussed above and are selected to be sufficiently broad so as to avoid biasing the inferred constraints. 
We also verified, in all main joint analyses performed in this work, 
that the resulting posterior distributions are not dominated by the 
priors, thanks to the sufficiently wide ranges adopted for each parameter.}

The statistical analysis was performed with the \texttt{GetDist} package\footnote{\url{https://github.com/cmbant/getdist}}, which we used to extract numerical results, including one-dimensional posterior distributions and two-dimensional marginalized probability contours. In what follows, we describe the likelihood functions and methodology adopted throughout this work.
\\

\begin{itemize}
\item \textit {Transverse Baryon Acoustic Oscillations}: 
We use transverse BAO measurements, $\{ \theta_{\rm BAO}(z) \}$, from a set of cosmological distance measures obtained through a methodology that weakly depends on fiducial cosmology~\citep{Sanchez11,Carnero12}. 
Our analysis employs 17 measurements of $\theta_{\rm BAO}$: 14 data points\footnote{We also account for the correlation between these measurements due to the overlap of their redshift bins. The corresponding covariance matrix is presented in Table~II of Ref.~\cite{Menote22}.} listed in Table~I of Ref.~\cite{Menote22}, covering the redshift range $0.35 \leq z \leq 0.63$, together with three additional independent measurements at $z = 2.225$~\cite{deCarvalho18}, $z = 0.11$~\cite{deCarvalho21}, and $z = 1.725$~\cite{Avila25}, the latter being presented in a companion work.

These 2D BAO measurements were obtained by identifying the transverse BAO signature in the two-point angular correlation function of pairs of cosmic objects, such as quasars or 
galaxies~\citep{Ribeiro2025, Sanchez11}. 
In this framework, the BAO signal emerges as an excess probability at a characteristic angular scale, reflecting the preferred separation imprinted by acoustic waves in the early Universe.  

For a successful detection of the transverse BAO signal, three key factors must be taken into account~\citep{deCarvalho20, Avila2024}: 
\begin{itemize}
    \item[(i)] the number density of the cosmic objects;
    \item[(ii)] the surveyed area on the sky; and
    \item[(iii)] precise redshift measurements, which are essential to accurately define the sample of cosmic objects within the thin redshift bin under analysis.
\end{itemize}

Once these observational factors are accounted for, a measurement of the BAO angular scale at a given redshift $z$ provides a direct estimate of the angular diameter distance through
\begin{equation}
D_A(z) = \frac{r_d}{(1+z)\,\theta_{\rm BAO}(z)} \,,
\end{equation}
where $r_d$ denotes the comoving sound horizon at the baryon drag epoch. It is therefore crucial to have a robust estimate of $r_d$ to accurately infer the function $D_A(z)$ from the measured $\theta_{\rm BAO}(z)$ data.  

{As already discussed in Section~\ref{sec:introduction}, it is important to emphasize that 2D BAO measurements do not provide new independent cosmological information beyond what is already probed by 3D BAO ($D_M$ and $D_H$). However, because 2D BAO are obtained through a distinct observational pipeline with minimal modeling assumptions and independent systematics, they serve as a valuable tool for consistency checks and cross-validation.}

In this work, we unify the previously described $\theta_{\rm BAO}$ measurements into a new compilation, providing a more comprehensive dataset for cosmological analyses.  It is worth emphasizing that this new compilation differs significantly from the one presented in~\citep{Nunes:2020hzy}, in terms of the included datasets and the methodology used to analyze the transverse BAO measurements. As a result, this novel compilation provides greater statistical leverage and robustness in constraining cosmological parameters, especially in high-redshift regimes and model-independent analyses. The resulting dataset of these measurements is commonly referred to as \texttt{2D~BAO}.

\item \textit{Baryon Acoustic Oscillations} (\textbf{DESI-DR2}): We employ BAO measurements from the second data release of the DESI survey, which include observations from galaxies, quasars~\cite{DESI2025}, and Lyman-$\alpha$ tracers~\cite{DESI:2025zpo}. These measurements, detailed in Table~IV of Ref.~\cite{DESI2025}, cover the effective redshift range $0.295 \leq z \leq 2.330$, divided into nine bins. The BAO constraints are expressed in terms of the transverse comoving distance $D_{\mathrm{M}}/r_d$, the Hubble distance $D_{\mathrm{H}}/r_d$, and the angle-averaged distance $D_{\mathrm{V}}/r_d$, all normalized to the comoving sound horizon at the drag epoch, $r_d$. We also take into account the correlation structure among these quantities through the cross-correlation coefficients $r_{V,M/H}$ and $r_{M,H}$, which describe the covariance between different BAO measurements. This dataset is referred to as \texttt{DESI-DR2}.

\item \textit{Cosmic Microwave Background}: We use the temperature and polarization anisotropy power spectra of the CMB measured by the \textit{Planck} satellite~\cite{Planck:2018vyg}, along with their cross-spectra from the 2018 legacy data release. Specifically, we adopt the high-$\ell$ \texttt{Plik} likelihood for TT ($30 \leq \ell \leq 2508$), TE, and EE ($30 \leq \ell \leq 1996$), as well as the low-$\ell$ TT-only ($2 \leq \ell \leq 29$) and EE-only ($2 \leq \ell \leq 29$) \texttt{SimAll} likelihoods~\cite{Planck:2019nip}. In addition, we include the CMB lensing reconstruction based on the temperature four-point correlation function~\cite{Planck:2018lbu}. This dataset is referred to as \texttt{CMB}.

\item \textit{Type Ia Supernovae} (\textbf{SN Ia}): We employ the following recent SN Ia samples:

\begin{enumerate}
    \item[(i)] \textbf{PantheonPlus}: The PantheonPlus (PP) sample~\cite{Brout:2022vxf} provides distance-modulus measurements from 1701 light curves of 1550 distinct SN Ia events, spanning the redshift range $0.01 \leq z \leq 2.26$. This dataset is referred to as \texttt{PP}.

    \item[(ii)] \textbf{Union~3.0}: The Union~3.0 compilation~\cite{Rubin:2023jdq} consists of 2087 SN Ia in the redshift range $0.001 < z < 2.260$, with 1363 objects overlapping with the PantheonPlus sample. This dataset, referred to as \texttt{Union3}, adopts a Bayesian hierarchical modeling approach to address systematic uncertainties and measurement errors. 
    This dataset is referred to as \texttt{Union3}. 

    \item[(iii)] \textbf{DESY5}: The Dark Energy Survey Year~5 (DESY5) sample~\cite{DES:2024jxu} includes 1635 photometrically classified SN Ia with redshifts $0.1 < z < 1.3$, along with 194 low-redshift SN Ia ($0.025 < z < 0.1$) shared with the PantheonPlus compilation. This dataset is referred to as \texttt{DESY5}.
\end{enumerate}

\end{itemize}

Finally, we note that throughout our analysis we employ state-of-the-art assumptions for 
Big Bang Nucleosynthesis (BBN), which is particularly sensitive to constraints on the 
physical baryon density, $\omega_{\rm b} = \Omega_{\rm b} h^2$. In particular, we use the 
\texttt{BBN} likelihood that incorporates the most precise measurements of primordial 
light element abundances currently available: the helium mass fraction, $Y_P$, as reported 
in~\cite{Aver:2015iza}, and the deuterium-to-hydrogen ratio, $y_{\rm DP} = 10^5\, n_D / n_H$, 
from~\cite{Cooke:2017cwo}. As is standard practice, BBN information is included in all 
analyses where CMB data are unavailable.

\section{Main Results}
\label{sec:results}
In this section, we present the main results of our analysis, derived from a series of statistical studies using the datasets introduced in the previous section. We discuss the constraints obtained for different cosmological models, highlight their implications, and compare them where relevant. A detailed discussion of the results, including potential systematic effects and consistency checks, follows.

\begin{figure*}[htbp!]
    \centering
    \includegraphics[width=0.46\textwidth]{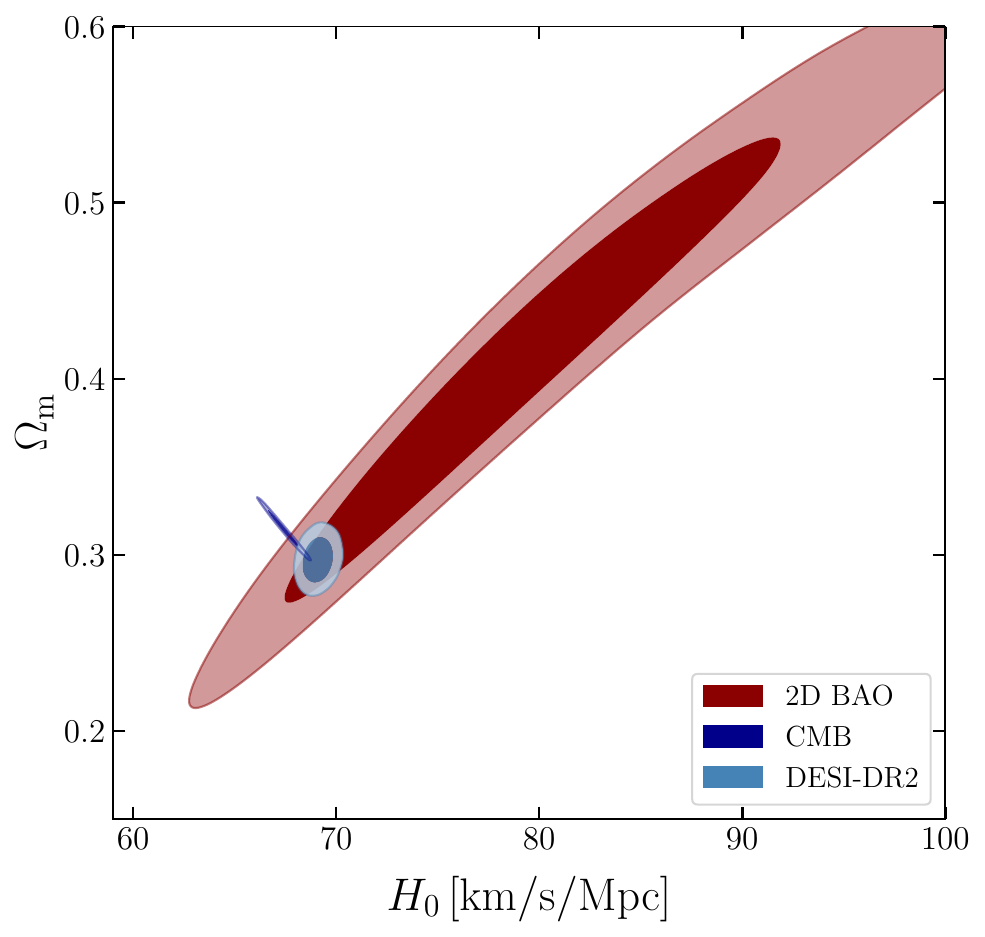}
    \hspace{0.03\textwidth}
    \includegraphics[width=0.4725\textwidth]{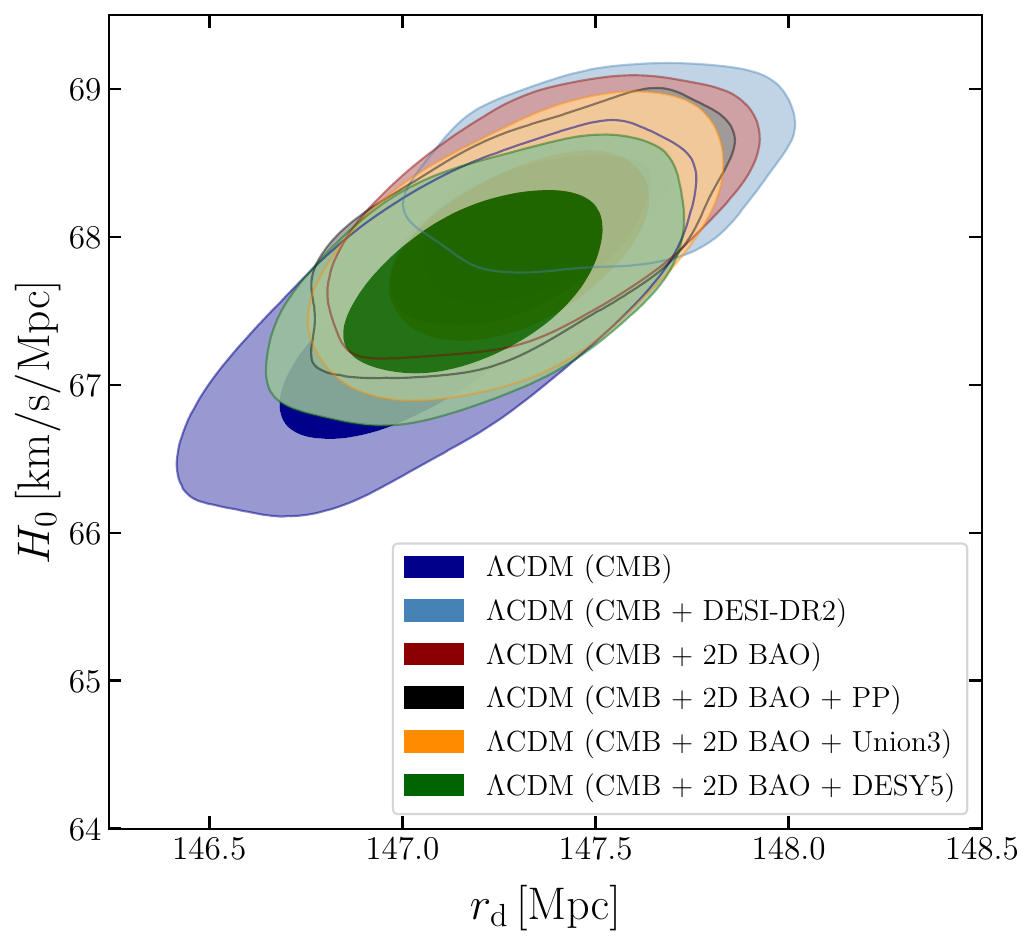}
    \caption{Left panel: Two-dimensional confidence contours at the 68\% and 95\% confidence levels for the parameters $\Omega_{\mathrm{m}}$ and $H_0$, obtained from different data combinations as indicated in the legend, within the framework of the $\Lambda$CDM model. Right panel: Same as in the left panel, but for several joint analyses combining CMB data with the 2D~BAO samples; the corresponding DESI-DR2 analyses are also shown in both panels to illustrate the impact of 2D~BAO on the constraining power while preserving consistency among the datasets.}
    \label{fig2}
\end{figure*}

\begin{table*}[htpb!]
\centering
\caption{Summary table of cosmological parameter constraints showing 68\% confidence levels (CL) for parameters of the $\Lambda$CDM model, obtained using our 2D BAO compilation and DESI-DR2 data in joint analyses with CMB measurements. We present both cases where the sum of neutrino masses is fixed and where it is allowed to vary freely, corresponding to the respective joint analyses. This allows a direct assessment of the impact of neutrino mass assumptions on the derived cosmological parameters. We quote 68\% confidence levels for all parameters, except for the neutrino mass in eV units, for which we report the 95\% upper limit.}
\renewcommand{\arraystretch}{1.55}
\resizebox{\linewidth}{!}{
\begin{tabular}{l||c|c|c|c|c} 
\hline
\textbf{Parameter} & \textbf{CMB} & \textbf{CMB + 2D BAO} & \textbf{CMB + DESI-DR2} & \textbf{CMB + 2D BAO + $\Sigma m_\nu$} & \textbf{CMB + DESI-DR2 + $\Sigma m_\nu$} \\
\hline \hline
$10^{2}\omega_{b}$ & $2.239\pm 0.015$ & $2.251\pm 0.013$ & $2.256^{+0.014}_{-0.012}$ & $2.250\pm 0.014$ & $2.254\pm 0.013$ \\
$\omega_{\rm cdm}$ & $0.1199\pm 0.0012$ & $0.11836\pm 0.00087$ & $0.11760\pm 0.00065$ & $0.11846\pm 0.00091$ & $0.11792\pm 0.00071$ \\
$100\,\theta_s$ & $1.04188\pm 0.00029$ & $1.04204\pm 0.00028$ & $1.04210\pm 0.00028$ & $1.04203\pm 0.00027$ & $1.04206\pm 0.00028$ \\
$\ln(10^{10}A_s)$ & $3.046\pm 0.015$ & $3.051^{+0.013}_{-0.015}$ & $3.055\pm 0.014$ & $3.049\pm 0.015$ & $3.051\pm 0.015$ \\
$n_s$ & $0.9659^{+0.0044}_{-0.0039}$ & $0.9695\pm 0.0037$ & $0.9714\pm 0.0035$ & $0.9694\pm 0.0036$ & $0.9708\pm 0.0035$ \\
$\tau_{\rm reio}$ & $0.0547\pm 0.0077$ & $0.0587^{+0.0064}_{-0.0079}$ & $0.0612\pm 0.0072$ & $0.0579^{+0.0070}_{-0.0082}$ & $0.0595\pm 0.0077$ \\
${\displaystyle \Sigma}\, m_{\nu}$ [eV] & -- & -- & -- & $< 0.0810$ & $< 0.0698$ \\
\hline
$H_0 \, [\text{km s}^{-1}\,\text{Mpc}^{-1}]$ & $67.45\pm 0.55$ & $68.16^{+0.41}_{-0.37}$ & $68.49\pm 0.29$ & $68.29\pm 0.47$ & $68.61\pm 0.31$ \\
$\sigma_8$ & $0.8113\pm 0.0060$ & $0.8091^{+0.0055}_{-0.0063}$ & $0.8083\pm 0.0059$ & $0.813^{+0.011}_{-0.0067}$ & $0.8149^{+0.0089}_{-0.0071}$ \\
$\Omega_{\rm m}$ & $0.3143\pm 0.0076$ & $0.3047^{+0.0048}_{-0.0054}$ & $0.3002\pm 0.0037$ & $0.3033\pm 0.0058$ & $0.2991\pm 0.0038$ \\
$r_d \, [\mathrm{Mpc}]$ & $147.09\pm 0.27$ & $147.36\pm 0.22$ & $147.52\pm 0.20$ & $147.36\pm 0.23$ & $147.46\pm 0.20$ \\
\hline \hline
\end{tabular}
}
\label{table_LCDM}
\end{table*}

\begin{table*}[htpb!]
\centering
\caption{Summary table of cosmological parameter constraints showing 68\% CL for parameters of the $\Lambda$CDM model, obtained from joint analyses of our 2D BAO compilation with different Type Ia supernova (SNIa) samples. These results highlight the impact of including SNIa data on the precision of cosmological parameters and the complementarity with 2D BAO measurements.}
\renewcommand{\arraystretch}{1.55}
\resizebox{\linewidth}{!}{
\begin{tabular}{l||c|c|c} 
\hline
\textbf{Parameter} & \textbf{CMB + 2D BAO + PP} & \textbf{CMB + 2D BAO + Union3} & \textbf{CMB + 2D BAO + DESY5} \\
\hline \hline
$10^{2}\omega_{b}$ & $2.248\pm 0.013$ & $2.247\pm 0.014$ & $2.243\pm 0.014$ \\
$\omega_{\rm cdm}$ & $0.11874\pm 0.00089$ & $0.11880\pm 0.00088$ & $0.11937\pm 0.00089$ \\
$100\,\theta_s$ & $1.04201\pm 0.00029$ & $1.04198^{+0.00029}_{-0.00025}$ & $1.04196\pm 0.00029$ \\
$\ln(10^{10}A_s)$ & $3.049^{+0.012}_{-0.016}$ & $3.046\pm 0.015$ & $3.047\pm 0.014$ \\
$n_s$ & $0.9687^{+0.0035}_{-0.0039}$ & $0.9686\pm 0.0037$ & $0.9672\pm 0.0036$ \\
$\tau_{\rm reio}$ & $0.0577^{+0.0059}_{-0.0080}$ & $0.0566\pm 0.0076$ & $0.0561^{+0.0068}_{-0.0076}$ \\
\hline
$H_0 \, [\text{km s}^{-1}\,\text{Mpc}^{-1}]$ & $67.98\pm 0.39$ & $67.93\pm 0.42$ & $67.69\pm 0.40$ \\
$\sigma_8$ & $0.8095\pm 0.0060$ & $0.8085\pm 0.0059$ & $0.8106\pm 0.0058$ \\
$\Omega_{\rm m}$ & $0.3070\pm 0.0052$ & $0.3076\pm 0.0054$ & $0.3109\pm 0.0054$ \\
$r_d \, [\mathrm{Mpc}]$ & $147.30\pm 0.23$ & $147.30\pm 0.22$ & $147.19\pm 0.22$ \\
\hline \hline
\end{tabular}
}
\label{table_LCDM_SN}
\end{table*}

\subsection{Constraints in $\Lambda$CDM}

We start by examining the constraints on the minimal $\Lambda$CDM parameters obtained from different datasets independently. The matter density parameter, $\Omega_{\rm m}$, is found to be $0.421^{+0.073}_{-0.10}$ from 2D BAO, while $\Omega_{\rm m}=0.2974 \pm 0.0086$ is obtained with DESI-DR2, and $\Omega_{\rm m}=0.3143 \pm 0.0076$ from the CMB. 
{Although the 2D BAO analysis yields significantly broader uncertainties, its constraints are consistent at the $1\sigma$ level with the more precise DESI-DR2 and CMB measurements. 
The larger uncertainties in the 2D BAO–only constraints are not driven by prior-dominated effects, but instead reflect the $\sim 10\%$ statistical errors inherent to 2D BAO measurements, in contrast to the $\sim 1\%$ precision achieved by 3D BAO data.} 
In comparison, DESI-DR2 and CMB provide substantially tighter constraints, highlighting the greater statistical power of these datasets.

Similarly, the derived combination $r_d h$ takes values of $r_d h=98.1^{+2.9}_{-2.6}$ Mpc for 2D BAO, $r_d h=101.56 \pm 0.73$ Mpc with DESI-DR2, and $r_d h=99.21 \pm 0.95$ Mpc from the CMB. 
While 2D BAO provides slightly lower central values and broader uncertainties, all results remain compatible within their respective errors. These comparisons highlight the complementarity of 2D BAO, DESI-DR2, and CMB datasets in constraining the standard $\Lambda$CDM model, establishing a robust baseline for analyses of extended cosmological scenarios in the following sections. 
{Therefore, from a statistical perspective, we found no significant tensions between these datasets.}

Figure~\ref{fig2} (left panel) shows the parameter space of $\Omega_{\rm m}$ versus $H_0$ with 1$\sigma$ and 2$\sigma$ confidence level (CL) regions for the analyses discussed above. The left panel illustrates that the 2D BAO error contours are highly degenerate. Consequently, when analyzed in isolation, 2D BAO measurements are insufficient to tightly constrain key cosmological parameters, including $H_0$, $r_d$, and $\Omega_{\rm m}$. {This clearly demonstrates the crucial role of combining 2D BAO measurements with complementary cosmological probes. 
In particular, the inclusion of early-time information from the CMB provides a well-calibrated anchor for the sound horizon scale, $r_d$, which is only weakly constrained by 2D BAO data alone. 
By fixing, or tightly constraining, $r_d$, the CMB effectively breaks the strong geometrical and dynamical degeneracies present in the $\Omega_{\rm m}$–$H_0$ parameter space inferred from the analysis of 2D BAO data. 
As a result, joint analyses substantially reduce the allowed parameter space and lead to significantly more precise, and robust, constraints on $\Omega_{\rm m}$, $H_0$, and related cosmological parameters.}

In this sense, we combine 2D BAO data with CMB measurements to break the parameter degeneracies present in the BAO-only analyses. The joint 2D BAO + CMB analysis significantly improves the constraints on key cosmological parameters. For example, the Hubble constant is found to be $H_0 = 68.16^{+0.41}_{-0.37}
\; \text{km s}^{-1}\,\text{Mpc}^{-1}$, compared to $H_0 = 67.45 \pm 0.55
\; \text{km s}^{-1}\,\text{Mpc}^{-1}$ from CMB alone and 
$H_0 = 68.49 \pm 0.29\; \text{km s}^{-1}\,\text{Mpc}^{-1}$ when combining CMB with DESI-DR2. 
Similarly, the matter density parameter tightens to $\Omega_{\rm m} = 0.3047^{+0.0048}_{-0.0054}$, compared with $\Omega_{\rm m} = 0.3143 \pm 0.0076$ from CMB only and $\Omega_{\rm m} = 0.3002 \pm 0.0037$ for CMB + DESI-DR2.

The combination also provides precise measurements of the sound horizon, with $r_d = 147.36 \pm 0.22$ Mpc for CMB + 2D BAO, and slightly higher values for CMB + DESI-DR2 ($147.52 \pm 0.20$ Mpc). Other parameters, such as $\sigma_8$ and the spectral index $n_s$, are also marginally shifted but remain consistent with the CMB-only constraints~\citep{Planck:2018vyg}. 

All main statistical results are summarized in Table~\ref{table_LCDM}. Comparing the combinations, we see that both CMB + 2D BAO and CMB + DESI-DR2 significantly improve parameter constraints relative to CMB alone. However, the combination with DESI-DR2 consistently provides tighter bounds. For instance, the Hubble constant is measured as $H_0 = 68.16^{+0.41}_{-0.37}\; \text{km s}^{-1}\,\text{Mpc}^{-1}$ for CMB + 2D BAO, while CMB + DESI-DR2 achieves $H_0 = 68.49 \pm 0.29\; \text{km s}^{-1}\,\text{Mpc}^{-1}$. 
These comparisons highlight that while 2D BAO data help break degeneracies and improve constraints beyond the CMB alone, the DESI-DR2 measurements provide even stronger links to the fundamental parameters, due to their higher statistical precision and coverage. 
Overall, the table illustrates the complementarity and relative constraining power of these different dataset combinations within the $\Lambda$CDM framework.

{Currently, several high-quality SNIa compilations are available, and they have consistently provided robust observational constraints on cosmological parameters. 
As standardized candles, SNIa act as a powerful late-time distance probe, directly mapping the expansion history of the Universe at low and intermediate redshifts. 
In this role, they are particularly sensitive to the late-time dynamics of cosmic acceleration and to parameters governing the background expansion, such as the matter density parameter $\Omega_{\rm m}$ and the properties of dark energy.}

{Moreover, SNIa effectively serve as a late-time anchor in cosmological analyses by providing precise relative distance measurements that tightly constrain the shape of the Hubble diagram. 
Although SNIa alone do not fix the absolute distance scale, and therefore cannot independently determine $H_0$ without external calibration, their strong sensitivity to the late-time expansion history makes them especially valuable when combined with other probes. 
In joint analyses, SNIa help break down the degeneracies between $\Omega_{\rm m}$, dark energy ($w_0$ and $w_a$), and curvature parameters, while complementing early-time anchors such as the CMB or intermediate probes like BAO. 
As a result, SNIa play a central role in connecting early- and late-Universe measurements and in providing stringent and internally consistent constraints on the cosmological model.}

Following the state of the art in cosmological analyses, we adopt multiple SNIa datasets, as defined in Section \ref{sec:data_method}, and incorporate them in a joint analysis with 2D BAO measurements. This combined approach allows us to exploit the complementary constraining power of SNIa luminosity distances and 2D BAO scales, improving parameter determinations beyond what is possible with either dataset alone.

Table~\ref{table_LCDM_SN} summarizes the main results in $\Lambda$CDM analyses. When using the Pantheon+ (PP) sample, the Hubble constant is measured as $H_0 = 67.98 \pm 0.39\; \text{km s}^{-1}\,\text{Mpc}^{-1}$, slightly higher than the value obtained with the Union3 compilation ($H_0 = 67.93 \pm 0.42\; \text{km s}^{-1}\,\text{Mpc}^{-1}$) and the DESY5 sample ($H_0 = 67.69 \pm 0.40\; \text{km s}^{-1}\,\text{Mpc}^{-1}$). Similarly, the matter density parameter, $\Omega_{\rm m}$, shows minor variations: $\Omega_{\rm m}=0.3070 \pm 0.0052$ for PP, $\Omega_{\rm m}=0.3076 \pm 0.0054$ for Union3, and $\Omega_{\rm m}=0.3109 \pm 0.0054$ for DESY5.

Other derived parameters, such as $\sigma_8$ and the sound horizon $r_d$, remain stable across the different SNIa datasets, with $\sigma_8$ ranging from $0.8085$ to $0.8106$ and 
$r_d \simeq 147.2 \,–\, 147.3$ Mpc. 
These results indicate that while the choice of SNIa compilation slightly shifts the central values, the overall constraints remain robust. The combination of 2D BAO with CMB and SNIa consistently improves the precision of cosmological parameters compared to CMB + 2D BAO alone, demonstrating the complementary role of SNIa data in breaking parameter degeneracies and refining measurements within the $\Lambda$CDM framework.

Examining the Hubble constant values in Table~\ref{table_LCDM_SN}, we note that all three joint analyses (CMB + 2D BAO + SNIa) yield $H_0$ values in the range $67.69 \,–\, 67.98\; \text{km s}^{-1}\,\text{Mpc}^{-1}$, which remain in close agreement with the Planck CMB measurements. These values are systematically lower than the local distance ladder determinations, such as those from the SH0ES collaboration \cite{Riess:2025chq} ($H_0 \sim 73\; \text{km s}^{-1}\,\text{Mpc}^{-1}$), highlighting the persistent tension between early- and late-Universe probes. Although the inclusion of different SNIa samples slightly shifts the central value — DESY5 leading to the lowest $H_0$ and Pantheon+ the highest — the changes are within the $1\sigma$ uncertainties and do not alleviate the tension. This emphasizes that, within the $\Lambda$CDM framework, the combined CMB + 2D BAO + SNIa datasets reinforce the lower $H_0$ values preferred by early-Universe measurements, underscoring the need for either new physics or systematic re-evaluations to reconcile the discrepancy.

\begin{table*}[htpb!]
\centering
\caption{Same as Table~\ref{table_LCDM}, but for the dynamical dark energy model $w_{0}w_{a}$CDM. We quote 68\% confidence levels for all parameters, except for the neutrino mass in eV units, for which we report the 95\% upper limit. The last row reports the difference in minimum chi-square value, $\Delta \chi^2_{\rm min} = \chi^2_{{\rm min} \, (w_0w_{\rm a}\mathrm{CDM})} - \chi^2_{{\rm min} \,(\Lambda\mathrm{CDM})}$, where negative values indicate a better fit of the $w_0w_{\rm a}$CDM model compared to the $\Lambda$CDM model.}
\renewcommand{\arraystretch}{1.55}
\resizebox{\linewidth}{!}{
\begin{tabular}{l||c|c|c|c|c} 
\hline
\textbf{Parameter} & \textbf{CMB} & \textbf{CMB + 2D BAO} & \textbf{CMB + DESI-DR2} & \textbf{CMB + 2D BAO + $\Sigma m_\nu$} & \textbf{CMB + DESI-DR2 + $\Sigma m_\nu$} \\
\hline \hline
$10^{2}\omega_{b}$ & $2.242\pm 0.016$ & $2.241\pm 0.014$ & $2.241\pm 0.014$ & $2.236\pm 0.015$ & $2.240\pm 0.014$ \\
$\omega_{\rm cdm}$ & $0.1194^{+0.0012}_{-0.0014}$ & $0.1199\pm 0.0011$ & $0.11965^{+0.00097}_{-0.00084}$ & $0.1199^{+0.0012}_{-0.0010}$ & $0.11951\pm 0.00094$ \\
$100\,\theta_s$ & $1.04194\pm 0.00030$ & $1.04189\pm 0.00028$ & $1.04191\pm 0.00028$ & $1.04291^{-0.00055}_{-0.0015}$ & $1.04194\pm 0.00030$ \\
$\ln(10^{10}A_s)$ & $3.041\pm 0.015$ & $3.043\pm 0.013$ & $3.042^{+0.013}_{-0.014}$ & $3.045\pm 0.015$ & $3.043\pm 0.015$ \\
$n_s$ & $0.9672\pm 0.0043$ & $0.9658\pm 0.0040$ & $0.9664\pm 0.0037$ & $0.9651\pm 0.0043$ & $0.9665\pm 0.0037$ \\
$\tau_{\rm reio}$ & $0.0534\pm 0.0077$ & $0.0537\pm 0.0069$ & $0.0535^{+0.0064}_{-0.0077}$ & $0.0546\pm 0.0076$ & $0.0541\pm 0.0077$ \\
$w_{0}$ & $-1.10\pm 0.43$ & $-0.91\pm 0.39$ & $-0.44\pm 0.21$ & $-0.78^{+0.29}_{-0.38}$ & $-0.42^{+0.24}_{-0.20}$ \\
$w_{\rm a}$ & $< 0.0362$ & $-0.6^{+1.3}_{-1.0}$ & $-1.67\pm 0.59$ & $-1.13^{+1.1}_{-0.96}$ & $-1.75^{+0.63}_{-0.74}$ \\
${\displaystyle \Sigma}\, m_{\nu}$ [eV] & -- & -- & -- & $< 0.372$ & $< 0.186$ \\
\hline
$H_0 \, [\text{km s}^{-1}\,\text{Mpc}^{-1}]$ & $76.6\pm 8.4$ & $69.1^{+3.4}_{-4.0}$ & $63.8^{+1.6}_{-2.1}$ & $68.1\pm 2.8$ & $63.8^{+1.7}_{-2.2}$ \\
$\sigma_8$ & $0.887^{+0.087}_{-0.077}$ & $0.827^{+0.030}_{-0.034}$ & $0.782^{+0.015}_{-0.017}$ & $0.807\pm 0.027$ & $0.777\pm 0.021$ \\
$\Omega_{\rm m}$ & $0.252^{+0.067}_{-0.076}$ & $0.302\pm 0.031$ & $0.351\pm 0.021$ & $0.312^{+0.021}_{-0.030}$ & $0.352\pm 0.022$ \\
$r_d \, [\mathrm{Mpc}]$ & $147.20\pm 0.28$ & $147.08\pm 0.26$ & $147.14\pm 0.22$ & $147.05^{+0.26}_{-0.29}$ & $147.17\pm 0.22$ \\
\hline
$\Delta \chi^2 _{\rm min}$ & $-4.38$ & $-2.62$ & $-9.20$ & $-1.44$ & $-2.90$ \\
\hline \hline
\end{tabular}
}
\label{table_w0waCDM}
\end{table*}

\begin{table*}[htpb!]
\centering
\caption{Same as Table~\ref{table_LCDM_SN}, but for the dynamical dark energy model $w_{0}w_{a}$CDM. The last row reports the difference in minimum chi-square value, $\Delta \chi^2_{\rm min} = \chi^2_{{\rm min} \, (w_0w_{\rm a}\mathrm{CDM})} - \chi^2_{{\rm min} \,(\Lambda\mathrm{CDM})}$, where negative values indicate a better fit of the $w_0w_{\rm a}$CDM model compared to the $\Lambda$CDM model.}
\renewcommand{\arraystretch}{1.55}
\resizebox{\linewidth}{!}{
\begin{tabular}{l||c|c|c} 
\hline
\textbf{Parameter} & \textbf{CMB + 2D BAO + PP} & \textbf{CMB + 2D BAO + Union3} & \textbf{CMB + 2D BAO + DESY5} \\
\hline \hline
$10^{2}\omega_{b}$ & $2.241^{+0.016}_{-0.014}$ & $2.240^{+0.013}_{-0.015}$ & $2.240\pm 0.015$ \\
$\omega_{\rm cdm}$ & $0.1196\pm 0.0011$ & $0.1197\pm 0.0011$ & $0.1198\pm 0.0011$ \\
$100\,\theta_s$ & $1.04189\pm 0.00030$ & $1.04188\pm 0.00029$ & $1.04189\pm 0.00029$ \\
$\ln(10^{10}A_s)$ & $3.043\pm 0.015$ & $3.040\pm 0.014$ & $3.040\pm 0.015$ \\
$n_s$ & $0.9665\pm 0.0040$ & $0.9664\pm 0.0039$ & $0.9662\pm 0.0040$ \\
$\tau_{\rm reio}$ & $0.0538\pm 0.0076$ & $0.0525\pm 0.0075$ & $0.0523\pm 0.0076$ \\
$w_{0}$ & $-0.834\pm 0.062$ & $-0.69\pm 0.10$ & $-0.672\pm 0.071$ \\
$w_{\rm a}$ & $-0.71\pm 0.26$ & $-1.13\pm 0.38$ & $-1.20^{+0.33}_{-0.28}$ \\
\hline
$H_0 \, [\text{km s}^{-1}\,\text{Mpc}^{-1}]$ & $68.03\pm 0.64$ & $66.86\pm 0.86$ & $66.81\pm 0.61$ \\
$\sigma_8$ & $0.817\pm 0.010$ & $0.808\pm 0.011$ & $0.8082\pm 0.0096$ \\
$\Omega_{\rm m}$ & $0.3084\pm 0.0060$ & $0.3195\pm 0.0085$ & $0.3200\pm 0.0063$ \\
$r_d \, [\mathrm{Mpc}]$ & $147.14\pm 0.25$ & $147.13\pm 0.25$ & $147.12\pm 0.25$ \\
\hline
$\Delta \chi^2 _{\rm min}$ & $-5.86$ & $-11.34$ & $-22.60$ \\
\hline \hline
\end{tabular}
}
\label{table_cosmology_params}
\end{table*}

\begin{figure*}[htbp!]
    \centering
    \includegraphics[width=0.46\textwidth]{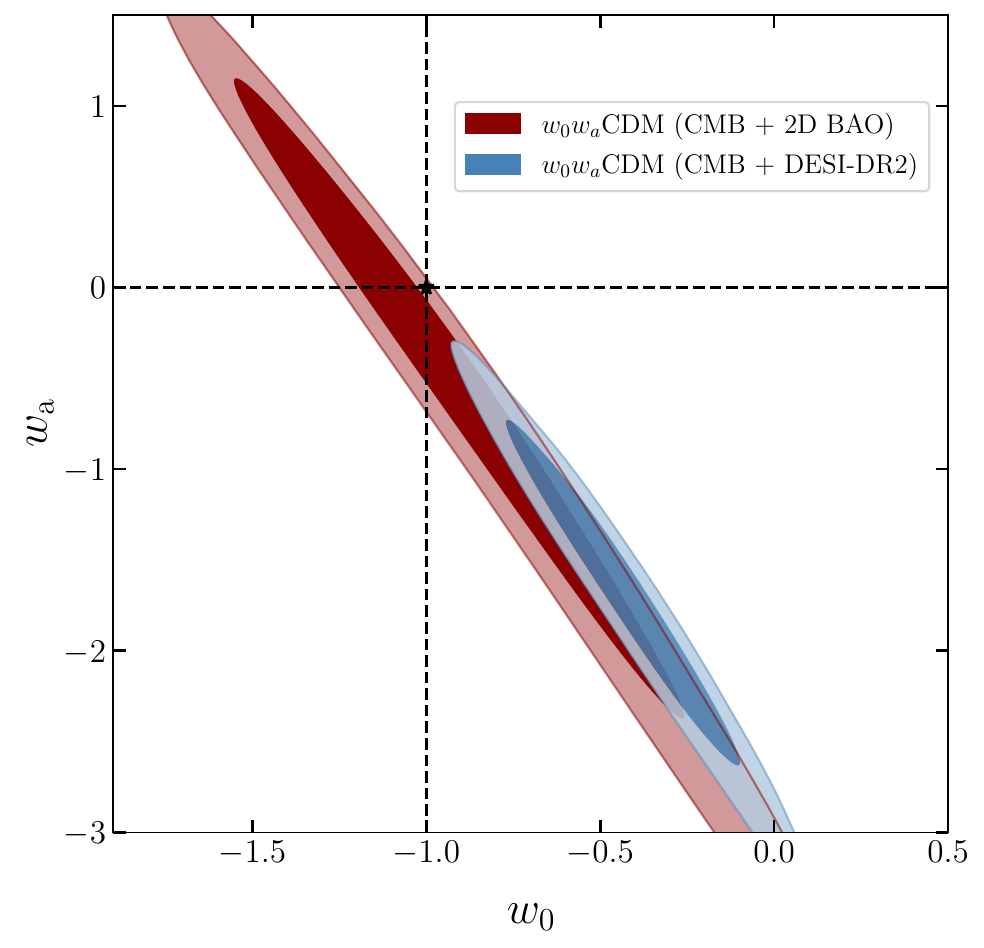}
    \hspace{0.03\textwidth}
    \includegraphics[width=0.465\textwidth]{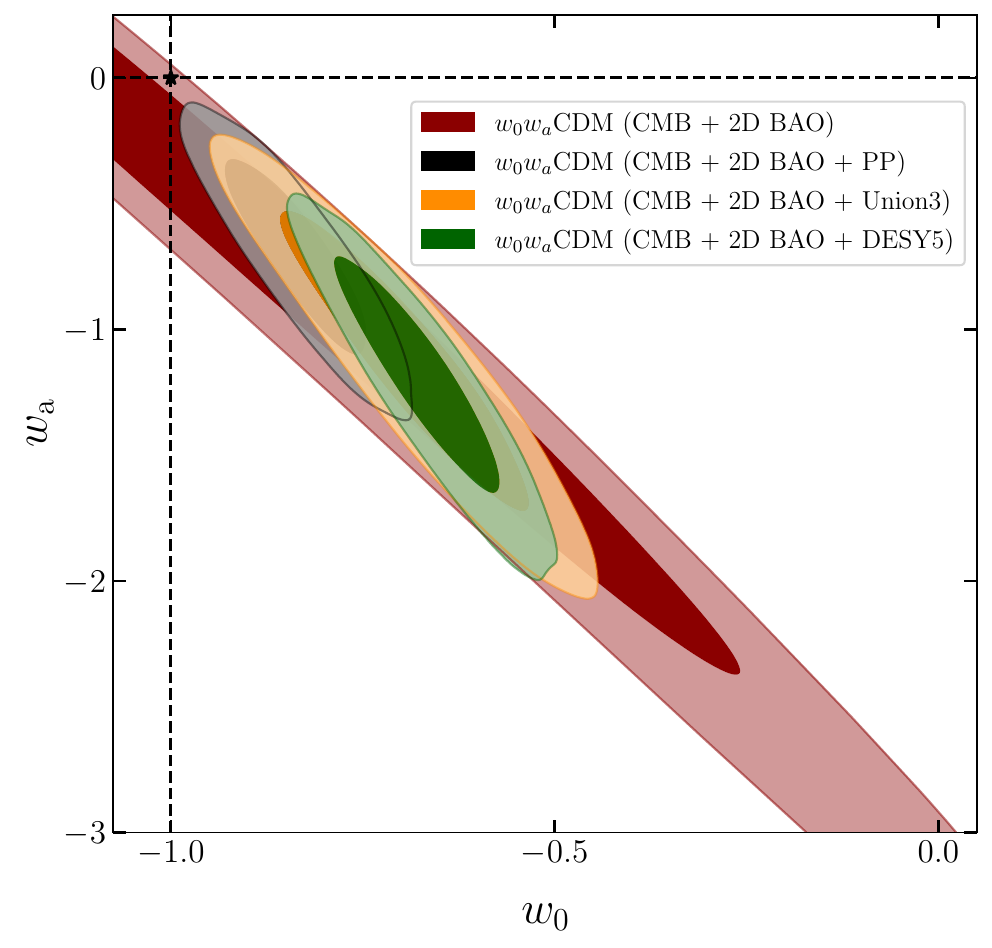}
    \caption{Left panel: Two-dimensional confidence contours at the 68\% and 95\% confidence levels for the parameters $w_0$ and $w_a$, obtained from different data combinations as indicated in the legend. Right panel: Same as in the left panel, but for several joint analyses combining CMB data with the 2D~BAO samples. For comparison, the DESI-DR2 analysis is included in the left panel to highlight both the consistency among the datasets and the differences in their constraining power.}
    \label{fig_w0waCDM}
\end{figure*}

{
Within the $\Lambda$CDM framework, our main results can be summarized as follows:}
\begin{itemize}
    \item {All dataset combinations yield statistically consistent and mutually compatible constraints on the cosmological parameters investigated.} 
    \item {When analyzed in isolation, 2D BAO measurements suffer from strong parameter degeneracies and therefore provide weak constraints.}
    \item {However, the combination of 2D BAO data with CMB data efficiently breaks these degeneracies through the early-time sound-horizon anchor, leading to significantly improved constraints on $H_0$, $\Omega_{\rm m}$, and $r_d$.}
    \item {The joint CMB + 2D BAO results are fully consistent with those obtained from CMB + DESI-DR2, which provides the highest statistical precision among the combinations considered.}
    \item {The inclusion of SNIa as late-time distance probes further tightens the constraints without inducing significant shifts in the central values, yielding $H_0 \simeq 67.7 - 68.0\;\mathrm{km\,s^{-1}\,Mpc^{-1}}$ and $\Omega_{\rm m} \simeq 0.31$.}
\end{itemize}

\subsection{Constraints in $w_0w_a$CDM}

Table~\ref{table_w0waCDM} summarizes the constraints on cosmological parameters for the $w_0 w_a$CDM model, obtained from a variety of dataset combinations. The analyses mirror those performed for the standard $\Lambda$CDM model, allowing for a direct comparison between the two scenarios. {However, unlike the $\Lambda$CDM case, we did not investigate constraints derived from 2D BAO measurements alone in the $w_0w_a$CDM framework.}

{This choice is motivated by the intrinsically limited constraining power of individual 2D BAO measurements on parameters governing the time evolution of dark energy. 
Owing to their relatively narrow redshift leverage, 2D BAO observables primarily probe the expansion history at discrete redshift slices and therefore exhibit only weak sensitivity to the dynamical parameter $w_a$, which encodes redshift-dependent departures from the cosmological constant case. 
As a result, analyses based solely on 2D BAO data lead to highly degenerate and poorly constrained parameter spaces in the $w_0$–$w_a$ plane, offering limited physical insight. 
For this reason, we restrict our $w_0w_a$CDM analyses to joint combinations in which 2D BAO data are complemented by datasets with broader redshift coverage and stronger sensitivity to late-time dynamics, such as CMB and SNIa data. 
In these joint analyses, 2D BAO measurements play a supporting role by anchoring the distance scale at intermediate redshifts, while the additional probes provide the necessary redshift leverage to constrain the time evolution of dark energy. 
This strategy ensures robust and physically meaningful constraints on $w_0$ and $w_a$, and avoids over-interpreting the limited information content of 2D BAO data when used in isolation.}

Focusing first on the dark energy equation-of-state parameters, we note that 
the constraints on $w_0$ and $w_a$ vary significantly depending on the dataset used. 
When using only CMB data, the uncertainties are relatively large ($w_0 = -1.10\pm0.43$, 
$w_a < 0.0362$), reflecting the limited ability of CMB alone to constrain dynamical 
dark energy. Including 2D BAO measurements or DESI-DR2 data substantially improves 
the constraints, driving $w_0$ closer to $-0.4$ to $-0.9$ and providing meaningful 
bounds on $w_a$. These trends demonstrate the importance of combining CMB with BAO data to break parameter degeneracies and tighten constraints. 
Although these shifts are not yet statistically significant enough to claim a detection of deviation, they indicate the sensitivity of combined datasets to departures from a pure cosmological constant model. 
Figure~\ref{fig_w0waCDM} (left panel) displays the parameter space in the $w_0$–$w_a$ plane for the CMB + 2D BAO and CMB + DESI DR2 combinations, allowing a direct comparison of their constraining power. As previously discussed, the CMB + 2D BAO data yield highly degenerate constraints, whereas the inclusion of DESI DR2 measurements significantly improves parameter localization. The resulting contours for CMB + DESI DR2 indicate a moderate statistical preference for the $w_0w_a$CDM model, with a confidence level exceeding $2\sigma$. This behavior, however, is consistent with several previous studies reported in the literature (e.g., \cite{Giare:2024gpk,DESI2025}).

The inferred value of the Hubble constant, $H_0$, shows a clear dependence on the choice of dataset. The CMB-only analysis yields a relatively high value, $H_0 = 76.6 \pm 8.4~\mathrm{km\,s^{-1}\,Mpc^{-1}}$, while the inclusion of 2D BAO data lowers both the central value and its associated uncertainty, reaching $H_0 \simeq 69.1^{+3.4}_{-4.0}~\mathrm{km\,s^{-1}\,Mpc^{-1}}$ for the CMB + 2D BAO combination.
A similar trend is observed for $\sigma_8$, indicating a better-constrained picture of structure formation when multiple probes are combined.

Other parameters, including the baryon and cold dark matter densities ($\omega_b$, $\omega_{\rm cdm}$), the sound horizon at the drag epoch $r_d$, and the scalar spectral 
index $n_s$, remain largely stable across the different dataset combinations, indicating that these quantities are already tightly constrained by the CMB data only.

\begin{figure*}[htbp!]
    \centering
    \includegraphics[width=0.7\linewidth]{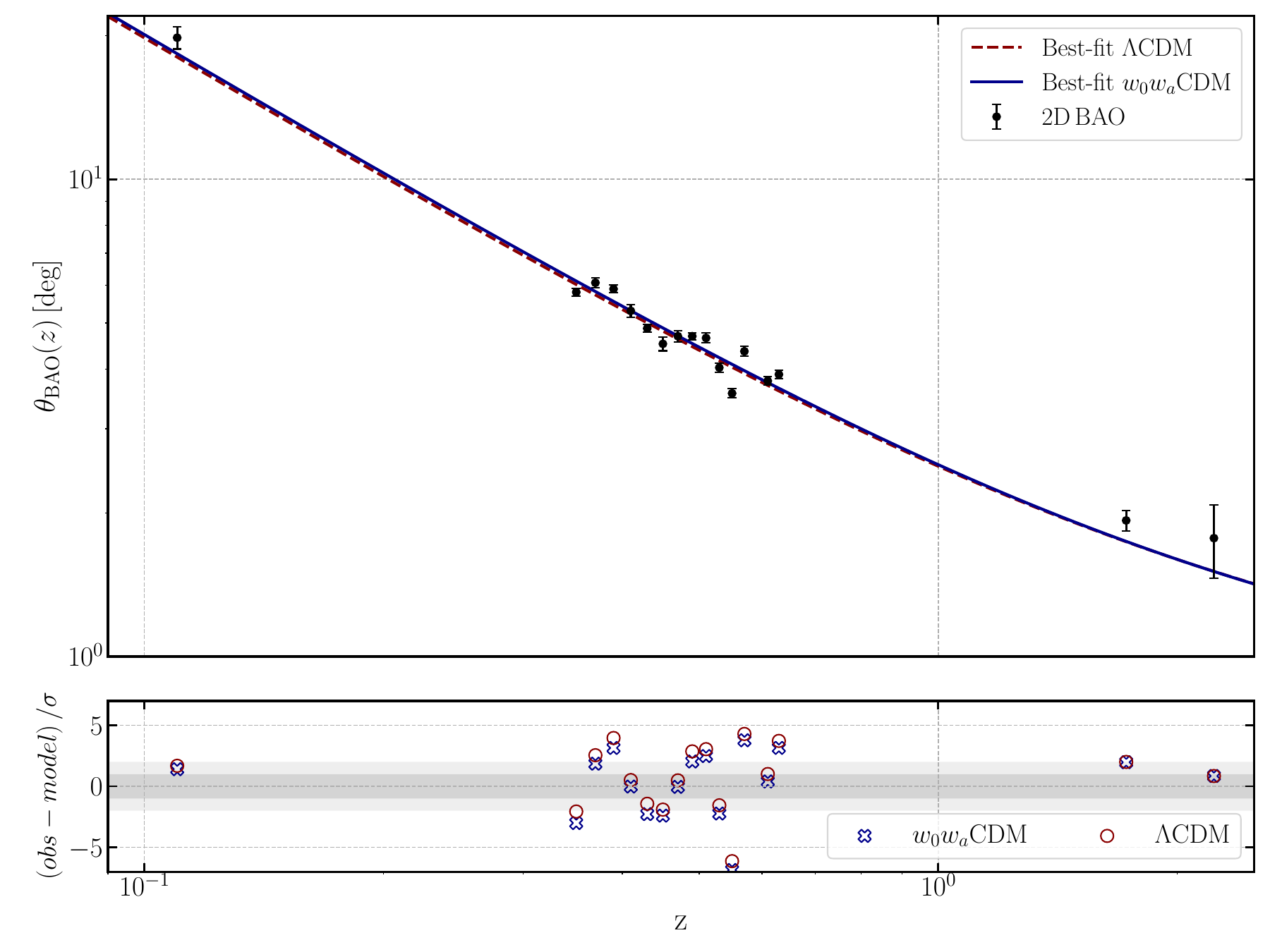}
    \caption{Best-fit models for the rescaled distance–redshift relation, $\theta_{\mathrm{BAO}}(z)$, in the $w_0w_a$CDM (solid line) and $\Lambda$CDM (dashed line) frameworks, derived from the combined analysis of CMB and DESI~DR2 data. The black dots correspond to the BAO measurement included in the study, as indicated in the legend. The vertical bars represent the $1\sigma$ observational uncertainties. 
\textit{Bottom panel:} Differences between the theoretical predictions and the observed BAO values, expressed in units of their respective observational errors.}
     \label{fig_thetaBAO}
\end{figure*}

Similar as done for $\Lambda$CDM model, Table~\ref{table_cosmology_params} summarizes the cosmological constraints obtained from the combination of CMB + 2D~BAO data with PP, the Union3 supernova sample, and the DES-Y5 dataset. 
In general, all combinations deliver mutually consistent results, indicating excellent internal concordance among current cosmological probes. 
The baryon and cold dark matter physical densities, $\omega_b$ and $\omega_{\rm cdm}$, are tightly constrained and show no significant dependence on the chosen dataset. 
The same holds for the sound horizon at the drag epoch, $r_d$, which remains stable around $147.1~\mathrm{Mpc}$ across all combinations. 
This confirms that the background cosmology at early times is robustly described within the standard framework, and any deviations from $\Lambda$CDM primarily arise from the late-time evolution of dark energy.

The dark energy equation-of-state parameters exhibit mild but coherent deviations 
from the cosmological constant scenario. In particular, we find 
$w_0 = -0.834 \pm 0.062$ and $w_a = -0.71 \pm 0.26$ for the 
CMB + 2D BAO + PP combination. When supernova datasets are included, the constraints shift toward more negative values of $w_a$, yielding 
$w_0 = -0.69 \pm 0.10$, $w_a = -1.13 \pm 0.38$ for 
CMB + 2D BAO + Union3, and 
$w_0 = -0.672 \pm 0.071$, $w_a = -1.20^{+0.33}_{-0.28}$ for 
CMB + 2D BAO + DES-Y5. 
This trend can be visually appreciated in the right panel of 
Figure~\ref{fig_w0waCDM} shows that the inclusion of SNe datasets slightly favors more dynamical dark energy models. 
For these joint analyses, we find the quantitative deviation from $\Lambda$CDM to be: \textbf{1.9\,$\sigma$} for CMB + 2D BAO + PP, \textbf{2.9\,$\sigma$} for CMB + 2D BAO + Union3, and \textbf{4.4\,$\sigma$} for CMB + 2D BAO + DESY5.

\noindent
For comparison, the DESI Collaboration (DR2) reports similar trends in their 
joint analyses:
\[
\begin{aligned}
w_0 &= -0.838 \pm 0.055, \quad w_a = -0.62^{+0.22}_{-0.19} 
&&, \\
w_0 &= -0.667 \pm 0.088, \quad w_a = -1.09^{+0.31}_{-0.27} 
&&, \\
w_0 &= -0.752 \pm 0.057, \quad w_a = -0.86^{+0.23}_{-0.20} 
&& ,
\end{aligned}
\]
from (CMB + DESI DR2 + PP), (CMB + DESI DR2 + Union3.0) and (CMB + DESI DR2 + DES-Y5), respectively.

\begin{figure}[htbp!]
    \centering
    \includegraphics[width=0.95\linewidth]{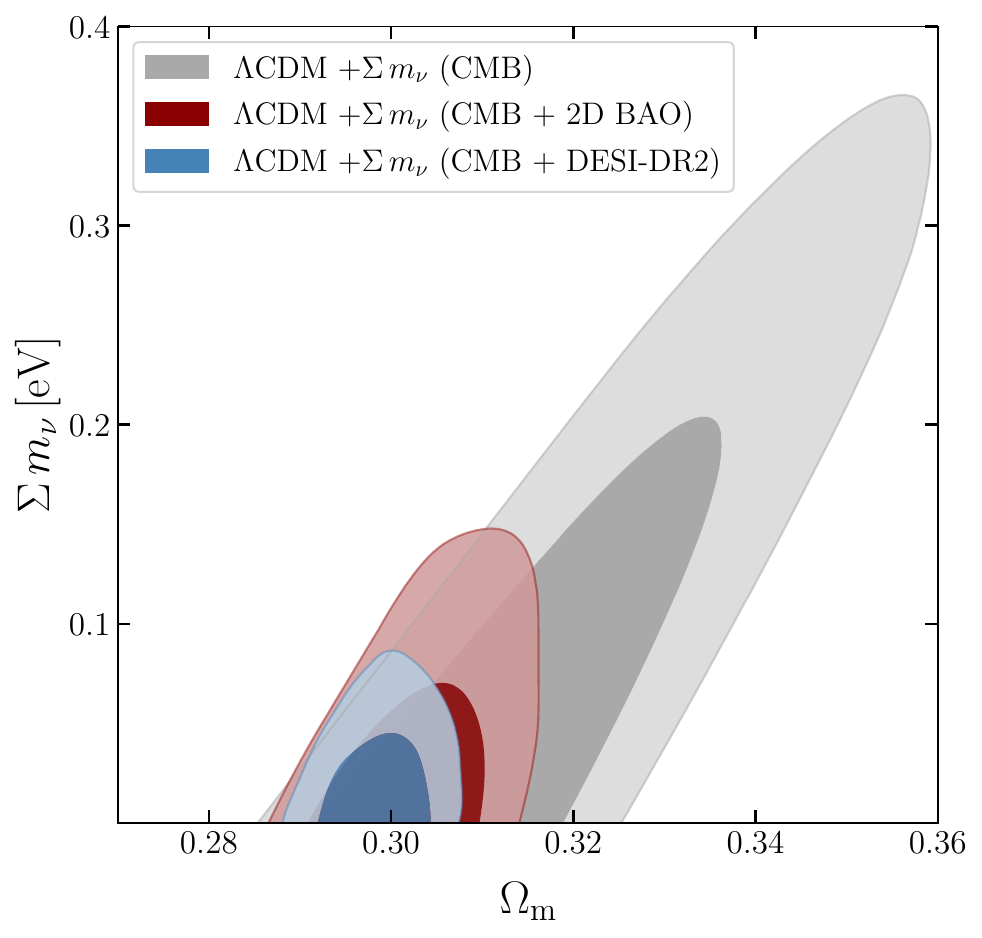}
    \caption{The 68\% and 95\% confidence regions for the sum of neutrino masses, $\Sigma m_{\nu}$, and the matter density parameter, $\Omega_{\rm m}$, within the $\Lambda$CDM framework. The analysis assumes a prior of $\Sigma m_{\nu} > 0~\mathrm{eV}$. CMB-only constraints reveal a strong positive correlation between $\Sigma m_{\nu}$ and $\Omega_{\rm m}$. Both 2D BAO and DESI BAO data, while largely insensitive to the neutrino mass, provide tight measurements of $\Omega_{\rm m}$, effectively breaking the geometric degeneracy and resulting in a more stringent upper limit on $\Sigma m_{\nu}$.}
    \label{fig_neutrinos_LCDM}
\end{figure}

\begin{figure*}[htbp!]
    \centering
    \includegraphics[width=0.95\textwidth]{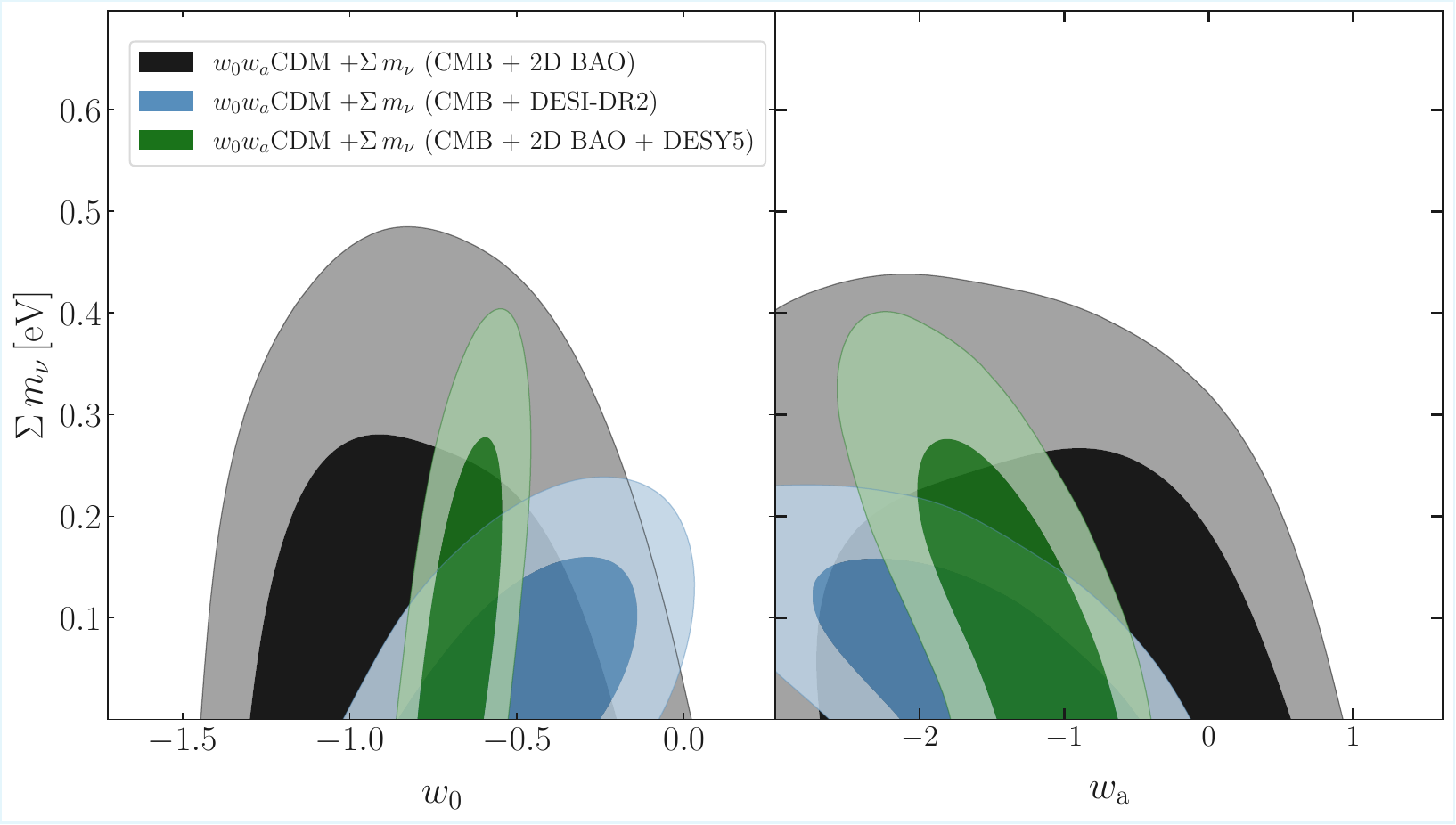}
    \caption{Limits on the sum of neutrino masses within the $w_0w_a$CDM framework. 
The shaded regions represent the 68\% and 95\% credible intervals of the posterior 
distribution, assuming a prior of $\Sigma m_{\nu} > 0~\mathrm{eV}$ in all cases.}
    \label{figw0waMnu}
\end{figure*}

Overall, our results are consistent with the DESI findings, reinforcing the indication of deviations from a pure cosmological constant behavior ($w_0=-1$, $w_a=0$). More importantly, from a comparative perspective, we can conclude that replacing the 2D~BAO compilation with DESI~DR2 leads to similarly robust observational constraints in joint analyses. {In other words, both datasets provide consistent and complementary information when combined with other cosmological probes.}

The derived Hubble constant also shows a modest dependence on the adopted 
supernova dataset. We obtain $H_0 = 68.0 \pm 0.6~\mathrm{km\,s^{-1}\,Mpc^{-1}}$ 
from the CMB + 2D BAO + PP combination, and slightly lower values of 
$H_0 \simeq 66.8~\mathrm{km\,s^{-1}\,Mpc^{-1}}$ when Union3 or DES-Y5 
are included. These results remain consistent with the Planck $\Lambda$CDM 
estimate and do not indicate any alleviation of the Hubble tension. 
Similarly, the amplitude of matter fluctuations, $\sigma_8$, and the total 
matter density, $\Omega_m$, exhibit stable and well-constrained values across 
all datasets, pointing toward a coherent picture of structure formation.

Figure~\ref{fig_thetaBAO} shows the theoretical prediction for the angular BAO scale, 
$\theta_{\mathrm{BAO}}$, in comparison with the corresponding observational measurements from the sample used in this work. The solid and dashed curves represent the best-fit predictions for the $\Lambda$CDM and $w_0w_a$CDM models, respectively. As can be seen, the theoretical curves provide an excellent description of the data across the entire redshift range. This comparison highlights the sensitivity of the 2D~BAO measurements to potential departures from the cosmological constant, thus providing a powerful test of dark energy dynamics.

{Within the $w_0w_a$CDM framework, the main results can be summarized as follows:}
\begin{itemize}
    \item {Constraints on the dark energy equation-of-state parameters, $w_0$ and $w_a$, are weak for CMB-only analyses, reflecting the limited sensitivity of early-time data to dynamical dark energy.}
    \item {The inclusion of BAO data, particularly DESI-DR2, significantly improves the constraints by breaking degeneracies, leading to a moderate ($\gtrsim 2\sigma$) statistical preference for dynamical dark energy relative to $\Lambda$CDM.}
    \item {When combined with CMB data, 2D~BAO measurements yield more degenerate constraints than DESI--DR2, but remain fully consistent with it, providing sensitivity to late-time deviations from a cosmological constant.}
    \item {The addition of Type Ia supernovae as late-time distance probes further tightens the constraints and systematically shifts $w_a$ toward negative values, with the strongest deviation from $\Lambda$CDM reaching $\sim 4\sigma$ for the CMB + 2D BAO + DES-Y5 combination.}
    \item {Other cosmological parameters, including $\omega_b$, $\omega_{\rm cdm}$, $r_d$, $\Omega_{\rm m}$, and $\sigma_8$, remain stable across all dataset combinations, indicating that deviations from $\Lambda$CDM are driven primarily by late-time dark energy dynamics rather than changes in early-Universe physics.}
\end{itemize}

{Although the $w_0w_a$CDM model allows for departures from a pure cosmological constant and exhibits a mild preference for a dynamical dark energy, the inferred values of $H_0$ remain consistent with CMB-based determinations and do not alleviate the Hubble tension.}

\subsection{Constraints on the Neutrino Mass}
\label{neutrino_results}

Cosmological observations offer one of the most promising avenues to constrain the absolute neutrino mass scale (see, e.g.,\cite{Lesgourgues:2006nd,Lesgourgues:2014zoa,Vagnozzi:2019utt,DiValentino:2024xsv,Escudero:2024uea} for a review), providing a complementary approach to laboratory measurements from oscillation and beta-decay experiments. This sensitivity arises because massive neutrinos leave subtle but measurable imprints on various cosmological observables, including the expansion history, the growth of large-scale structure~\citep{Marques2019, Marques20}, and the suppression of power on small scales. Nevertheless, the precise determination of the total neutrino mass depends critically on the cosmological model assumed in the analysis. State-of-the-art datasets have been extensively employed to investigate neutrino properties, particularly using BAO measurements from DESI DR2, as discussed in, e.g., \cite{Jiang:2024viw,Giare:2025ath,Wang:2025ker,RoyChoudhury:2025dhe,RoyChoudhury:2024wri,Elbers:2025vlz,Du:2025xes,Gariazzo:2022ahe}.

In our baseline framework, as outlined in the previous subsection, we adopt a total mass for the three active neutrino species fixed at the minimum value allowed by neutrino oscillation experiments, $\sum m_\nu = 0.06$ eV, consistent with the normal mass ordering scenario. In this fiducial configuration, the neutrino sector is modeled as one massive and two effectively massless states, which is adequate for most standard cosmological analyses.

To explore possible deviations and assess the robustness of our results, we extend our parameter space by allowing $\sum m_\nu$ to vary freely. Throughout this section, we adopt a model in which the three neutrino mass eigenstates are assumed to be degenerate in mass. This approximation reproduces with high accuracy the cosmological effects of both the normal and inverted mass orderings, and it has been shown to recover the correct total mass value in the case of a positive detection, without introducing significant bias~\cite{Lesgourgues:2006nd}. A uniform physical prior $\sum m_\nu > 0$ eV is imposed, ensuring the exploration of physically meaningful regions of parameter space. 

In this subsection, we present updated results obtained using the 2D~BAO compilation, combined with several complementary datasets discussed previously. We emphasize that these results are new in their own right, independent of the main objective, which is to analyze the updated 2D~BAO dataset.
Tables~\ref{table_LCDM} and \ref{table_w0waCDM} summarize our constraints on the neutrino mass for the $\Lambda$CDM and $w_0w_a$CDM models, respectively.

Allowing the sum of neutrino masses, $\Sigma m_{\nu}$, to vary freely leads to tight upper limits from our main joint analyses. 
Using the combination CMB + 2D~BAO, we obtain 
$\Sigma m_{\nu} < 0.081~\mathrm{eV}$ (95\%~CL), while the 
CMB + DESI~DR2 analysis yields an even stronger constraint of 
$\Sigma m_{\nu} < 0.0698~\mathrm{eV}$ (95\%~CL). 
The inclusion of large-scale-structure information from either 2D~BAO or DESI~DR2 
significantly enhances the sensitivity to $\Sigma m_{\nu}$ compared to CMB-only analyses, mainly by breaking degeneracies between the neutrino mass and matter density parameters. The slightly tighter bound obtained with DESI~DR2 reflects its higher precision in measuring the distance–redshift relation, confirming that both datasets provide robust and complementary constraints on the total neutrino mass. 
Importantly, the inclusion of the 2D BAO compilation delivers reliable constraints, demonstrating that this dataset constitutes an independent probe for testing neutrino mass effects in cosmology.

Figure~\ref{fig_neutrinos_LCDM} displays the parameter space in the 
$\Omega_{\rm m}$--$\Sigma m_{\nu}$ plane for the CMB, 
CMB + 2D~BAO, and CMB + DESI~DR2 analyses. 
It is evident that the inclusion of BAO data significantly tightens 
the upper limits on $\Sigma m_{\nu}$, with the strongest constraints obtained 
when either 2D~BAO or DESI~DR2 measurements are added. 
As discussed above for other cosmological parameters, we observe a similar trend 
for the neutrino mass: the DESI~DR2 dataset provides more stringent bounds, 
reflecting the higher precision of its distance–redshift measurements. 
This comparison highlights the complementary role of different BAO compilations 
and demonstrates that both 2D~BAO and DESI~DR2 contribute robust and independent 
constraints on the total neutrino mass.

\textit{Neutrino mass constraints in the $w_0w_a$CDM model.} 
To robustly constrain the sum of neutrino masses, $\Sigma m_{\nu}$, we combine 
CMB and 2D~BAO measurements with three supernova samples. 
Allowing $\Sigma m_{\nu}$ to vary freely results in upper limits that are somewhat 
weaker than those obtained in the $\Lambda$CDM case, reflecting the additional 
freedom introduced by the dynamical dark energy parameters. 
From the joint analyses, we find $\Sigma m_{\nu} < 0.290~\mathrm{eV}$ (95\%~CL) 
for CMB + 2D~BAO + PP, while including supernova datasets slightly relaxes the 
constraint: $\Sigma m_{\nu} < 0.323~\mathrm{eV}$ for CMB + 2D~BAO + Union3 
and $\Sigma m_{\nu} < 0.332~\mathrm{eV}$ for CMB + 2D~BAO + DES-Y5. 
This trend is expected, as the additional dark energy parameters ($w_0$, $w_a$) 
partially degenerate with $\Sigma m_{\nu}$, reducing the constraining power of the data. 
Nevertheless, all limits remain well below $0.35~\mathrm{eV}$, demonstrating that the combination of CMB, BAO, and supernova measurements continues to provide a robust 
upper bound on the total neutrino mass even in the context of dynamical dark energy models.

Figure~\ref{figw0waMnu} shows the parameter space of $\Sigma m_{\nu}$ versus $w_0$ 
and $\Sigma m_{\nu}$ versus $w_a$ for several of the joint analyses discussed above. These plots illustrate the statistical correlations between the neutrino mass and the dark energy parameters, as well as how different combinations of datasets constrain 
$\Sigma m_{\nu}$ within this extended parameter space.

In this subsection, we show that combining 2D BAO measurements with CMB data and other cosmological distance probes, such as SNIa, provides robust constraints on the neutrino mass. {Our analyses show that these constraints are consistent with those obtained from DESI~DR2 measurements. 
Accordingly, 2D~BAO can be regarded as a complementary probe for studying neutrino properties.}

{Within the context of cosmological neutrino mass constraints, our main results can be summarized as follows:}
\begin{itemize}
    \item {By allowing the total neutrino mass, $\Sigma m_{\nu}$, to vary freely, we obtain tight upper limits when BAO information is combined with CMB data, demonstrating the strong sensitivity of distance probes to neutrino mass effects.}
    \item {In the $\Lambda$CDM framework, we obtain $\Sigma m_{\nu} < 0.081~\mathrm{eV}$ (95\% CL) from CMB + 2D BAO and a slightly stronger bound of $\Sigma m_{\nu} < 0.0698~\mathrm{eV}$ (95\% CL) from CMB + DESI DR2, reflecting the higher statistical precision of the latter.}
    \item {The inclusion of either 2D BAO or DESI DR2 measurements significantly improves neutrino mass constraints relative to CMB-only analyses by breaking degeneracies between $\Sigma m_{\nu}$ and the matter density $\Omega_{\rm m}$.}
    \item {In the extended $w_0w_a$CDM model, the bounds on $\Sigma m_{\nu}$ are relaxed due to degeneracies with the dark energy parameters, but remain robust, with all joint analyses yielding $\Sigma m_{\nu} \lesssim 0.33~\mathrm{eV}$ at 95\% CL.}
    \item {The 2D BAO compilation delivers consistent constraints, confirming its role as an independent and complementary probe of neutrino mass effects in cosmology.}
\end{itemize}

\section{Final Remarks}
\label{sec:conclusions}

In this work, we have presented updated cosmological constraints using the 
2D~BAO compilation in combination with complementary datasets, including CMB, 
BAO DESI~DR2, and various supernova samples, considering both the $\Lambda$CDM and 
$w_0w_a$CDM frameworks. For the $\Lambda$CDM model, the combination of CMB and 2D~BAO data provides robust constraints on standard cosmological parameters and a stringent upper limit on the sum of neutrino masses, $\Sigma m_{\nu} < 0.081~\mathrm{eV}$ (95\%~CL). 
Incorporating DESI~DR2 measurements further tightens this bound to $\Sigma m_{\nu} < 0.070~\mathrm{eV}$, highlighting the high precision of DESI in constraining the matter density parameter and breaking the geometric degeneracy 
present in CMB-only analyses. In the context of dynamical dark energy models, $w_0w_a$CDM, allowing the dark energy equation-of-state parameters to vary introduces additional degeneracies with $\Sigma m_{\nu}$, leading to slightly weaker upper limits: $\Sigma m_{\nu} < 0.290~\mathrm{eV}$ for CMB + 2D~BAO + PP, and $\Sigma m_{\nu} < 0.332~\mathrm{eV}$ when including supernova datasets. Nevertheless, these constraints remain robust, demonstrating that both 2D BAO and DESI DR2 provide consistent and complementary probes of neutrino mass, even within the extended parameter space of dynamical dark energy.

Ultimately, our results show that current BAO measurements, particularly when combined with CMB and supernova data, provide powerful constraints on cosmological parameters, including the sum of neutrino masses. Both 2D~BAO and DESI~DR2 datasets yield mutually consistent and complementary results, reinforcing their crucial role in precision cosmology and in testing extensions beyond the minimal $\Lambda$CDM scenario.

In addition to the results presented here, we have demonstrated that this new 2D~BAO compilation is consistent with other leading cosmological probes. In particular, we have shown that 2D~BAO measurements can be combined with CMB-Planck data and multiple Type~Ia supernova samples without introducing significant tensions, reinforcing their reliability as an independent geometric probe.

{Looking ahead, 2D~BAO measurements are expected to play an increasingly important role in next-generation cosmological surveys. 
Upcoming public galaxy and quasar catalogs from 
DESI~\citep{DESI2024a, DESI2024b} 
and Euclid~\citep{Euclid2025} 
will enable high-precision determinations of the angular BAO scale, $\theta_{\rm BAO}$, over a wide range of redshifts. In this context, 2D~BAO observables will provide a powerful and largely model-independent cross-check of standard 3D~BAO analyses, offering an alternative avenue to assess systematic effects related to redshift-space distortions, galaxy bias, and assumptions about the fiducial cosmology. As such, 2D~BAO measurements will serve as a valuable complementary tool for validating cosmological constraints in the era of precision cosmology, and will represent an important resource for the broader community in testing the robustness of future cosmological results, especially when supported by independent pipelines acting as systematic cross-checks in upcoming high-precision surveys.}

\section*{Data Availability}
The datasets and products used in this research, including Boltzmann codes and likelihoods, will be made available upon reasonable request to the corresponding author following the publication of this article.


\begin{acknowledgments}
{We thank the referee for the careful reading of the manuscript and for the constructive comments and suggestions, which helped to significantly improve the clarity and robustness of this work.}
M.A.S. acknowledges support from CAPES and expresses gratitude to the Observatório Nacional for their hospitality during the development of this work. R.C.N. thanks the financial support from the Conselho Nacional de Desenvolvimento Científico e Tecnológico (CNPq, National Council for Scientific and Technological Development) under the project No. 304306/2022-3, and the Fundação de Amparo à Pesquisa do Estado do RS (FAPERGS, Research Support Foundation of the State of RS) for partial financial support under the project No. 23/2551-0000848-3. FA thanks to Fundação Carlos Chagas Filho de Amparo à Pesquisa do Estado do Rio de Janeiro (FAPERJ), Processo SEI-260003/001221/2025, for the financial support. AB acknowledges a CNPq fellowship. 
This work was carried out using computational resources provided by the Data Processing Center of the Observatório Nacional (CPDON).

\end{acknowledgments}

\bibliographystyle{apsrev4-1}
%
\end{document}